\documentclass[cits]{PoS}
\usepackage[authoryear,square]{natbib}
\bibpunct{(}{)}{;}{a}{}{,}

\title{Gravitational wave astronomy with the SKA}

\ShortTitle{GW astronomy}

\author{
\speaker{G.\,H.\,Janssen}$^1$, 
G.\,Hobbs$^2$,
M.\,McLaughlin$^3$,
C.\,G.\,Bassa$^1$, 
A.\,T.\,Deller$^1$,
M.\,Kramer$^{4,5}$,
K.\,J.\,Lee$^{6,4}$, 
C.\,M.\,F.\,Mingarelli$^{4,7,8}$,
P.\,A.\,Rosado$^{9,10}$,
S.\,Sanidas$^5$,
A.\,Sesana$^{11}$,
L.\,Shao$^{12}$, 
I.\,H.\,Stairs$^{13}$, 
B.\,W.\,Stappers$^5$, 
J.\,P.\,W.\,Verbiest$^{14,4}$ \\
\scriptsize{
$^1$Netherlands Institute for Radio Astronomy (ASTRON), The Netherlands; E-mail: \email{janssen@astron.nl};
$^2$CSIRO Astronomy and Space Science, Australia;
$^3$Department of Physics, West Virginia University, USA;
$^4$Max-Planck-Institut f\"ur Radioastronomie, Bonn, Germany;
$^5$University of Manchester, UK;
$^6$Kavli Institute for Astronomy and Astrophysics, Peking University, P.R. China;
$^7$Theoretical Astrophysics, California Institute of Technology, USA;
$^8$University of Birmingham, UK;
$^9$Max-Planck-Institut f\"ur Gravitationsphysik, AEI Hanover, Germany;
$^{10}$Swinburne University of Technology, Australia
$^{11}$Max-Planck-Institut f\"ur Gravitationsphysik, AEI Golm, Germany;
$^{12}$School of Physics, Peking University, P.R. China;
$^{13}$University of British Columbia, Canada;
$^{14}$Universit\"at Bielefeld, Germany
}}

\abstract{On a time scale of years to decades, gravitational wave (GW)
  astronomy will become a reality.  Low frequency
  ($\sim$10$^{-9}$ Hz) GWs are detectable through long-term
  timing observations of the most stable pulsars.
  Radio observatories worldwide are currently carrying out observing
  programmes to detect GWs, with data sets being shared through
  the International Pulsar Timing Array project.
One of the most likely sources of low frequency GWs 
are supermassive black hole binaries (SMBHBs), detectable as a
background due to a large number of binaries, or as continuous or
burst emission from individual sources.
No GW signal has yet been detected, but stringent constraints are
already being placed on galaxy evolution models.
The SKA will bring this research to fruition.  

In this chapter, we describe how timing observations using SKA1 will
contribute to detecting GWs, or can confirm a detection if a first
signal already has been identified when SKA1 commences
observations. We describe how SKA observations will identify the
source(s) of a GW signal, search for anisotropies in the background,
improve models of galaxy evolution, test theories of gravity, and
characterise the early inspiral phase of a SMBHB system.

We describe the impact of the large number of millisecond pulsars to
be discovered by the SKA; and the observing cadence, observation
durations, and instrumentation required to reach the necessary
sensitivity.  We describe the noise processes that will influence
the achievable precision with the SKA.
We assume a long-term timing programme using the \skamid\ array and
consider the implications of modifications to the current design. We
describe the possible benefits from observations using \skalow.
Finally, we describe GW detection prospects with SKA1 and SKA2, and
end with a description of the expectations of GW astronomy.
}

\FullConference{
Advancing Astrophysics with the Square Kilometre Array\\
June 8-13, 2014\\
Giardini Naxos, Italy}

\newcommand\skamid{SKA1-MID}
\newcommand\skalow{SKA1-LOW}

\begin{document}

\section{Introduction}

Radio observations of binary pulsars provide irrefutable evidence that
gravitational waves (GWs) exist. The impact of their emission on e.g. the
orbital period of binary pulsars can be used to test general
relativity and other theories of gravity
\citep[e.g.]{tw82,kbc+04,ckl+04}. The effect of external GWs on pulsar
timing \citep{det79} can be used to make a direct detection of
low-frequency GWs through the analysis of a large ensemble of pulsars,
forming a galactic-scale GW detector: a Pulsar Timing Array
\citep[PTA;][]{hd83,rom89,fb90}.  Global efforts are underway to
directly detect low-frequency GWs via PTA experiments.

In this chapter we describe how a PTA project for the SKA will allow
the characterisation of GW signals.  PTA projects are based on pulsar
timing observations in which pulse times of arrival (ToAs) for a
sample of pulsars are measured.  The ToAs are subsequently compared
with predictions for those arrival times based on simple physical
models of the pulsars.  Any differences between the predicted and
measured ToAs are referred to as the ``timing residuals'' and indicate
that a physical phenomenon, not (or incorrectly) included in the
model, is affecting the pulse ToAs at a measurable level.
Residuals caused by the irregular rotation of a given pulsar,
variability in the interstellar medium, or inaccuracies in the timing
model parameters are typically different between pulsars.  In
contrast, GWs passing the Earth will lead to timing residuals whose
functional form will depend upon the pulsar-Earth-GW angle. PTA
projects therefore differ from more traditional pulsar timing
experiments in that they aim to extract common signals present within
the timing residuals for multiple pulsars in order to detect GWs and
other correlated signals, like fluctuations in global atomic time
standards \citep{hcm+12} or unmodelled effects in the Solar system
ephemeris that is used to convert pulse ToAs to the Solar system
barycentre \citep{chm+10}.

Long-term observations of millisecond pulsars were carried out at
telescopes around the world starting in the 1980s and 1990s, producing
data sets that can be usefully added to modern PTA observations
\citep[see e.g.][]{ktr94}.  Three PTAs are presently in existence.
The European Pulsar Timing Array \citep[EPTA; see][for a review]{kc13}
uses five radio telescopes in Europe (the 100-m Effelsberg telescope
in Germany, the 76-m Lovell telescope at Jodrell Bank in the UK, the
94-m equivalent Nan\c cay telescope in France, the 64-m Sardinia radio
telescope in Italy, the 94-m equivalent Westerbork Synthesis Radio
Telescope in The Netherlands) and complementary low-frequency
($<250$\,MHz) observations obtained with LOFAR \citep{vwg+13,sha+11}
for some pulsars. The North American Nanohertz Observatory for
Gravitational Waves \citep[NANOGrav;][]{dfg+13}, uses the 305-m
Arecibo and 100-m Green Bank telescopes, and the Parkes Pulsar Timing
Array \citep[PPTA;][]{mhb+13}, uses the 64-m Parkes telescope. All
collaborations have well established pulsar timing programmes on at least 20
millisecond pulsars with time baselines of 10 years or more. Regular
timing observations for these pulsars are being obtained over a wide
range of observing frequencies (300 to 3000\,MHz) to mitigate
the effects of variations in dispersion. Since 2008, the three PTAs are
collaborating as members of the International Pulsar Timing Array
\citep[IPTA;][]{man13}.
Over the years, all PTAs have increased their sensitivity to GWs
through improvements in instrumentation, software, and observing
cadence and data span.  Furthermore, the EPTA is using interferometric
techniques to form the LEAP telescope, a tied-array of the five large
European radio telescopes (\citealt{ks10}; Bassa et al.  in prep.). In
the near future it is expected that PTA observations will be also
carried out using the SKA Pathfinder telescope MeerKAT, and with the
Five-hundred-meter Aperture Spherical Telescope
\citep[FAST;][]{hdm+14}.

There is a broad GW spectrum, rich with information ranging from
processes in the early Universe to compact binary
coalescences. 

\begin{figure}\centering
\includegraphics[width=1.0\textwidth]{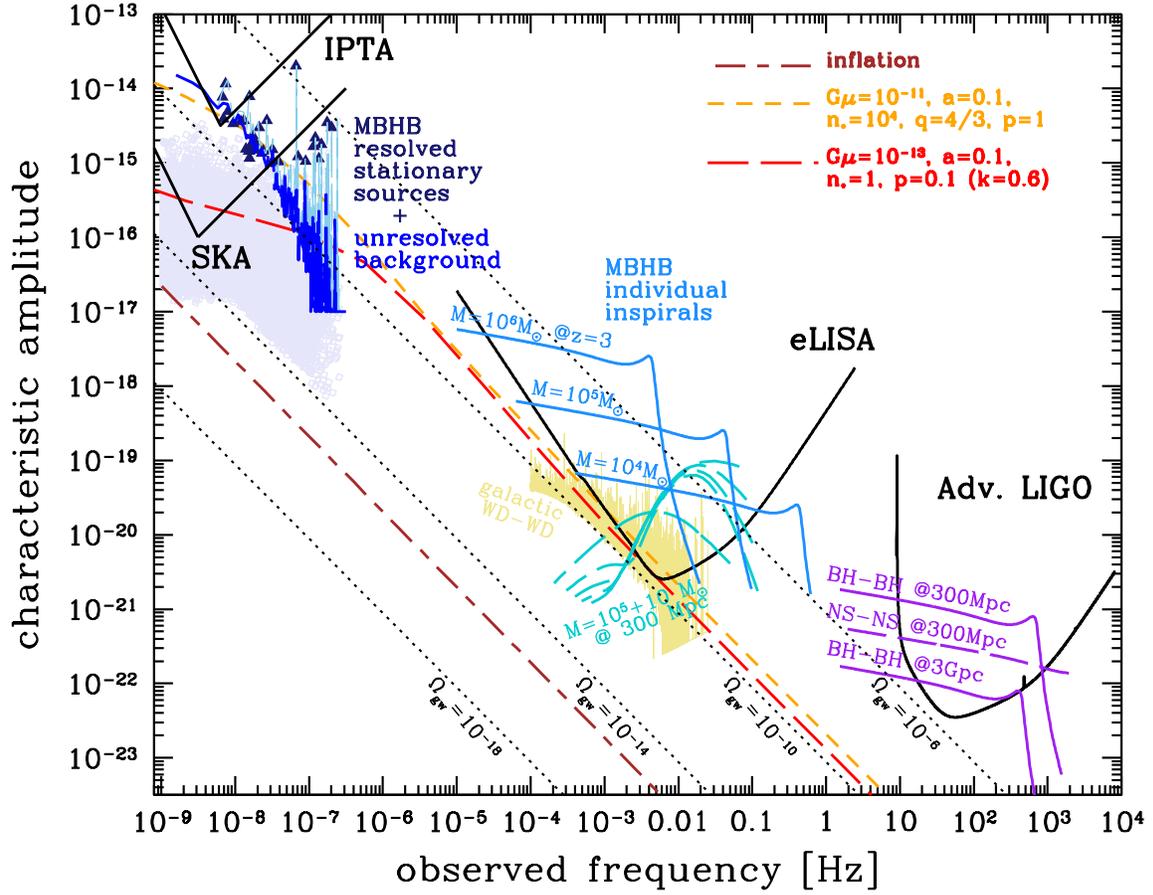} 
\caption{The gravitational wave landscape: characteristic amplitude
  ($h_c$), vs frequency.  In the nHz frequency range a selected
  realisation of the expected GW signal from the cosmological
  population of SMBHBs is shown. Small lavender squares are individual
  SMBHB contributions to the signal, the dark blue triangles are loud,
  individually resolvable systems and the blue jagged line is the
  level of the unresolved background. Nominal sensitivity levels for
  the IPTA and SKA are also shown. In the mHz frequency range, the
  eLISA sensitivity curve is shown together with typical circular SMBHB
  inspirals at z=3 (pale blue), the overall signal from Galactic WD-WD
  binaries (yellow) and an example of extreme mass ratio inspiral
  (aquamarine, only the first 5 harmonics are shown). In the kHz range
  an advanced LIGO curve (based on calculations for a single
  interferometer) is shown together with selected compact object
  inspirals (purple). The brown, red and orange lines running through
  the whole frequency range are expected cosmological backgrounds from
  standard inflation and selected string models, as labeled in
  figure. Black dotted lines mark different levels of GW energy
  density content as a function of frequency ($\Omega_{gw} \propto
  h_c^2f^2$). }\label{fg:gwsource}
\end{figure}

Cosmological GW signatures at frequencies of $\sim
10^{-16}\,$Hz are targeted by cosmic microwave
background (CMB) polarisation experiments such as BICEP2 and Planck \citep[e.g.][]{bicep,planck}.
Ground-based interferometer GW observatories, like the
(advanced) LIGO and Virgo, and a space-based mission, like the
proposed eLISA, operate at $\sim1$--$10^3$\,Hz and
$\sim10^{-4}$--$1$\,Hz, respectively \citep[see e.g.][for overviews]{prrh11,eLisa}.  Therefore, PTAs with sensitivities peaking
at frequencies of $10^{-9}$--$10^{-7}$\,Hz represent a unique
complementary laboratory to other GW experiments and are able to
detect and study unique classes of sources (see
Figure\,\ref{fg:gwsource}).

Throughout this chapter we concentrate on the detection and subsequent
analysis of GWs.  However, the data sets produced for this work will
be useful in numerous ways; including studying the interstellar
medium, the pulsars themselves, terrestrial time standards, planetary
ephemerides, the solar wind, and testing theories of gravity.  We
first describe the expected sources of GW emission detectable by PTAs
(\S\ref{sec:sources}) .  In \S\ref{sec:studyGWs} we describe how the
SKA will enable both GW detection and GW source characterisation. In
\S\ref{sec:noise} we highlight various limiting noise processes that
need to be accounted for in the detection analyses. In
\S\ref{sec:realistic} we describe a realistic PTA for the SKA which
includes a plausible observing strategy. The effect of a cut in the
baseline design is described in \S\ref{sec:cut}. We describe the
future of GW astronomy with the SKA in \S\ref{sec:gwastronomy}.

\section{Sources of gravitational waves}\label{sec:sources}

There are numerous GW sources that could lead to detectable signatures
in pulsar data sets.  Most recent work has concentrated upon searching
for a GW background formed by the superposition of a large number of
GWs from supermassive black hole binaries \citep[SMBHBs; see
  e.g.][]{svc08,rwh+12,oms10}, cosmic strings \citep[e.g.][]{sbs12} or
inflation \citep[e.g.][]{tzz+14}. Various algorithms have been
published to constrain such a background or to attempt to detect the
background \citep[see e.g.][]{jhv+06,ych+11,vlj+11,sbs12,dfg+13,
  src+13}.  These bounds have assumed that the GW background spectrum
is a power law defined by an amplitude $A$ and spectral index
$\alpha$.  
The stochastic GW backgrounds of the sources probed by PTAs have a
range of spectral indices, depending on the details of the theoretical
models, with the most widely used values being $\alpha \sim -2/3$ for
SMBHBs without environment interactions \citep{phi01}, $\alpha \sim
-7/6$ for high emission modes from cosmic string cusps and $\alpha
\sim -1$ for slow-roll inflation.
Current bounds are beginning to rule out some extreme models for the
coalescence rate of SMBHBs and for the tension and loop size of cosmic
strings.  The most stringent published limit, from PPTA data, on
characteristic amplitude due to a stochastic GW background is
$2.4\times10^{-15}$ \citep[at a reference frequency of
  yr$^{-1}$;][]{src+13}.  Figure \ref{fg:gwsource} shows a nominal
upper bound at the level we expect to reach with the full IPTA ($h_c
\sim 4 \times 10^{-15}$ at $5\times 10^{-9}$Hz, which corresponds to
about $h_c \sim 10^{-15}$ at a frequency of yr$^{-1}$ for an
extrapolated $f^{-2/3}$ power law) , and the expected levels to be
reached by the SKA ($h_c \sim 10^{-16}-10^{-17}$ at a reference
frequency of yr$^{-1}$).

Pulsar timing can also be used to search for individual sources of
GWs: individual SMBHBs that produce a continuous wave.  A single
(effectively non-evolving) circular binary system of supermassive
black holes produces a sinusoidal signal.  Various algorithms have
recently been developed to search for such waves.  These include a
frequentist-based approach \citep{jllw04,yhj+10} and Bayesian methods
\citep{vlml09,lwk+11,bs12,esc12,pbsd13,ell13,abb14}.  Pulsar timing
experiments are sensitive to sources emitting GWs with periods from a
few weeks (defined by the average observing cadence in the PTA) to
around the total duration of the data set. Using the algorithms
mentioned above, bounds have been placed on such GWs in this frequency
range.  For existing PTAs, GWs with periods of one or half a year are
undetectable because of the necessity of fitting for each pulsar's
position and parallax during the analysis (this can be improved by
using astrometric parameters independently measured through VLBI
observations; see \S\ref{sec:vlbi}).  Figure\,\ref{fg:gwsource} shows
examples of individually detectable SMBHB systems in a selected
synthetic realisation of the cosmic SMBHB population using dark-blue
triangle symbols. Another source of GWs in the PTA band is caused by
the merger of two supermassive black holes; this may lead to a
permanent deformation of their surrounding space-time, which can be
detectable as a memory event that takes the form of a ``glitch'' in
the timing residuals \citep{vl10,cj12}.

\section{Detecting gravitational waves}\label{sec:studyGWs}

Direct gravitational wave detection and its subsequent confirmation is
likely to be a long process.  Currently, searches for GWs with PTAs
are setting evermore stringent upper limits on the amplitude of the
stochastic GW background.  At some time (our predictions suggest that
this will be soon), the signatures of the GWs (most likely red noise
from a GW background) will become apparent in the timing residuals of
the pulsars with the lowest rms timing residuals. At this point the
bounds will not improve with time and it will become possible to study
the signal, determine its amplitude, frequency (for a continuous wave
source) or spectral index (for a background). However, confirmation of
the GW nature of the signal will only come when the expected
correlation signature for an isotropic stochastic GW background
(Hellings \& Downs curve; \citeyear{hd83}), is found in the timing
residuals of many pulsars.  Making a highly significant measurement of
such correlations requires much longer data sets than is needed for a
bound.  We therefore consider the following three scenarios for the
SKA PTA project: 1) no sign of a GW signal has been seen in the
existing PTA data, 2) a red noise signal has been identified, but the
data sets are insufficient to ``make a detection'' and 3) a
GW detection has already been made.  We conclude this section by
highlighting the role of Very Long Baseline Interferometry for
improving the SKA's sensitivity to GWs.

\subsection{No signal has been seen}

With the current limits \citep[e.g.][]{src+13} starting to constrain
the low-redshift SMBHB population, it is likely that, by the time the
SKA1 is commissioned, data from existing PTA experiments will already
indicate the presence of a GW signal.  However, if no such signal is
observed then it is likely that either the pulse arrival times have
not been measured with sufficient precision, that jitter noise is
dominating the data sets, that not enough pulsars are being observed
or that timing noise is affecting the detectability of the signal.
The SKA can provide significant improvements in all these areas.  With
its unprecedented sensitivity the SKA should be able to observe a
large number of currently unusable millisecond pulsars.  The timing
precision achievable for many pulsars will be limited by fluctuations
in the pulse shape from individual pulses.  This ``jitter noise'' (see
\S\ref{sec:noise} for more details) can be mitigated using the SKA by
forming sub-arrays, and thereby integrating for longer on multiple
pulsars in parallel.  With a much larger number of pulsars, timed with
better precision than currently available, it is likely that GW
detection can be made with shorter data spans (Rosado, Sesana \& Gair in
prep.). Having a shorter time span may also mitigate issues relating
to timing noise.  Moreover, we will be able to combine new data with
archival data collected up to SKA operation, so that the crucial time
baseline can be extended, albeit with a lower sensitivity.

\subsection{Red noise signal has already been identified}

If red noise (or a low-frequency signal) has been identified then the
SKA data sets can be used in conjunction with the longer existing data
sets to improve their sensitivity to GWs.  Research is ongoing to make
reasonable predictions for GW detection for a specified real (or
future) data set.  \cite{sejr13} made an initial attempt to calculate
the scaling of the signal-to-noise of an idealised PTAs as the signal
builds up in the array. They demonstrate that the significance of a
detection of a GW signal will increase with longer data spans at a
different rate for these three different regimes: 1) where the GW
signal is weak and the white noise dominates, 2) where the GW signal
is strong and 3) an intermediate regime where only the power in the
lowest frequencies of the GW background is above the white noise
level.  A realistic array is likely to be complex and contain some
pulsars for which the GW signal is weak compared to other red noise
signals in the data, and other pulsars for which the signal will be
comparatively strong.  Consequently, it is challenging to make
analytic predictions for the time to detection for a given array.  A
full analysis will require an improved understanding of the expected
signal strength and the noise processes (such as timing noise and
jitter noise), as well as a model-independent characterisation of the
statistical properties of the signal.

\subsection{GW detection has already been made}\label{sec:detectionDone}

If a GW signal has already been detected and published then the SKA
would begin the exciting work of studying the properties of the GWs
and characterising the GW background and sources (see also
\S\ref{sec:gwastronomy}).  For instance an SKA PTA would:

\begin{itemize}
\item \emph{Confirm the detection:} When a stochastic GW background
  has been detected, the signficance of the detection will increase as
  the quality and durations of data sets increase. SKA data sets would
  provide dramatically increased sensitivity and could be used to
  provide a confirmation of the detection, with a shorter timespan of
  data, different instrumentation, and different sample of millisecond
  pulsars.
\item \emph{Confirm the nature of the GW signal:} The induced
  residuals from a background of cosmic strings, the inflationary era,
  and merging supermassive black holes will all have the same
  characteristic angular signature. Futhermore, they all have red
  power spectra which will initially be indistinguishable. The SKA
  should be able to distinguish between the different spectral indices
  for the various sources.  A confirmed GW background from
  supermassive black hole mergers would provide strong evidence of the
  existence for a large population of SMBHBs where each binary is
  separated by less than a parsec, proving another crucial prediction
  of the hierarchical model of structure formation
  \citep{ses13a}. Moreover, \cite{msmv13} show that some level of
  anisotropy may be present in the stochastic GW background, and
  characterisation of the GW energy density at different angular
  scales carries information on the nature of the GW background.  With
  increasing spatial and frequency resolution, SKA will be also able
  to identify the loudest astrophysical GW sources, disentangling
  individual SMBHBs from the overall GW background \citep[e.g.][]{bs12,pbsd13}
\item \emph{Detect a turn-over in the GW spectrum:} For a background
  of SMBHBs, the expected simple power-law form for the
  characteristic strain spectrum is only valid for circular binaries
  whose evolution is driven solely by GW emission
  \citep{ses13b,rws+14}.  Any spectral turn-over would indicate
  interactions between the supermassive black holes and their environment
  (for instance, the effects of stellar scattering the presence of
  circumbinary discs). Furthermore, the GW spectrum for a background
  formed from cosmic strings could also have a complicated shape in
  the region of interest for PTA experiments \citep{sbs13}.

\end{itemize}

\subsection{The role of VLBI}\label{sec:vlbi}

For each pulsar in a PTA, the effect of a gravitational wave can be
partially absorbed into parameters in the pulsar timing model,
reducing low-frequency sensitivity.  It is therefore advantageous to obtain
independent measurements of the parameters that can be held fixed in
the pulsar timing model.  
The most common method of providing external information is via Very
Long Baseline Interferometry (VLBI), which yields very accurate
positions in the International Celestial Reference Frame (ICRF). By
repeating such measurements over time, pulsar reference positions,
proper motions, parallaxes, and in some cases orbital parameters are
obtained \citep[e.g.][]{dbl+13}.  In addition to being an independent
source of information, unaffected by low-frequency GWs, VLBI astronomy
gives higher precision than is possible via timing in almost all cases
\citep[see][for more details]{pgr+14}. A caveat is that at present
transferral from the ICRF to the Solar system barycentric frame
reduces the usefulness of the VLBI-derived \emph{reference position}
values, as the uncertainty in the frame tie (presently
$\sim$milliarcseconds) is orders of magnitude larger than the formal
uncertainty (but see however the third point below).

The three main benefits of VLBI information for gravitational wave
searches are as follows:

\begin{enumerate}
\item The sensitivity is greatly improved to GWs with
  periods of $\sim$0.5 years and $\sim$1.0 years, due to the
  GW contribution no longer being absorbed into the
  fit for parallax and proper motion (plus reference position if
  transferrable; e.g. \citealt{mcc13}).
\item{If very accurate distances can be derived for one or more
  pulsars, such that the remaining uncertainty is a small fraction of
  a gravitational wavelength, then sensitivity to individual sources
  can be greatly enhanced.  In this case, the contribution to the
  timing residuals from the gravitational wave at the pulsar (the
  ``pulsar term'') can be coherently accounted for \citep{lwk+11},
  which is a particular benefit for the pulsars with the highest
  timing precision.  If several such measurements can be made and
  multiple individual gravitational wave sources can be detected, then
  it will even be possible to determine the distances of other pulsars
  in the array to high precision using the gravitational wave fit
  directly \citep{lwk+11,bp12}}
\item{By comparing a large sample of VLBI positions and pulsar timing
  positions, it will be possible to improve the planetary ephemeris
  used to convert pulsar arrival times to the solar system barycentre,
  which will lead to improved accuracy for timing-derived pulsar
  positions \citep{mcc13}.}
\end{enumerate}

\section{Limiting noise processes}\label{sec:noise} 

The raw sensitivity of the SKA will enable ToAs to be determined with
much higher precision and will allow a much larger sample of pulsars
to be included in PTA experiments.  However, even though the ToA
timing precision will increase, a limit may exist in timing precision
due to various effects like pulse jitter on short time scales,
intrinsic pulsar timing noise on longer time scales and interstellar
medium effects such as scattering \citep{cs12}.
In this section we describe the various noise processes that, if
uncorrectable, will limit the achievable precision with the SKA.

\subsection{Pulse jitter}\label{sec:jitter}

The phenomenon of ``jitter noise'' results from intrinsic variability
in the shape of individual pulses from a pulsar. This has been shown
to be the limiting factor in the achievable timing precision for
PSR\,J0437$-$4715 \citep{ovh+11}, where a floor of the timing precision
of $\sim40$\,ns in a one-hour observation (and worse with shorter
observations) is reached, even when using the largest telescopes
available. Similar results have been found for PSR\,J1713+0747 using
Arecibo observations \citep{sc12}.

Jitter noise is thought to be, at some level, a limiting noise
process. However, an improvement of nearly 40\% in the rms timing
residual for PSR\,J0437$-$4715 was achieved by using information from
the polarised pulse profile \citep{ovdb13}, with the result ultimately
limited by variable Faraday rotation in the Earth's
ionosphere. Current studies of jitter noise are limited by the
sensitivity of available telescopes, with only a handful of
millisecond pulsars having detectable single pulses. As telescopes
become larger, we will be able to significantly improve our
understanding of the jitter phenomenon.  This, along with robust
calibration of data, could possibly lead to the development of
strategies to remove the effects of jitter from SKA data.

If jitter noise is uncorrectable, the SKA holds no
advantage for a jitter-dominated pulsar over a less-sensitive
telescope.  It is, however, likely that the SKA will observe many
pulsars that are in the jitter-dominated regime.  In order to maximise
the scientific output it will therefore be essential to sub-array the
telescope, enabling multiple pulsars to be observed simultaneously for
long sessions.

\subsection{Timing noise}

An analysis of timing irregularities for 366 pulsars allowed the first
large-scale analysis of timing noise over time-scales of $>$10\,yr
\citep{hlk10}.  The youngest pulsars were shown to be dominated by the
recovery from glitch events (sudden increases in the spin frequency),
while the timing irregularities for older pulsars seemed to exhibit
quasi-periodic structure.  Millisecond pulsars were included in the
sample, but the data set used (from the Lovell telescope at Jodrell
Bank) did not allow an analysis of timing noise with the precision
necessary for PTA-research.  The first major assessment of timing
noise on the precision timing of millisecond pulsars concluded that
timing noise is present in most millisecond pulsars and will be
measurable in many objects when observed over long data spans
\citep{sc10}.

There is currently no proven method for removing timing noise.
Furthermore, the power spectrum of timing noise is red,
with most of the power at low frequencies, and for some MSPs can be even 
steeper than the expected spectrum of the stochastic GW
background. Timing noise will therefore provide a stringent limit for
long-term timing projects.  The SKA will need to either observe
pulsars that exhibit small amounts of timing noise, observe a
sufficiently large number of pulsars to achieve the science goals
before timing noise dominates the residuals, or observe over long
enough data spans for the timing noise level to plateau.

For young pulsars, timing noise can be modelled as a process in which
the spin-down rate of the pulsar flips between two stable states
\citep{lhk+10}. These variations are accompanied by pulse profile
changes as well.  This leads to the possibilities that 1) millisecond
pulsar timing noise can also be modelled as a two-state process and
that 2) the state at a given time can be determined from the pulse
shape.  If true, it may be possible to completely remove the effect of
timing noise.

\subsection{Effects from the interstellar medium}\label{sec:dm}

The free electron content of the Galaxy affects radio wave propagation
in a number of different ways. Most importantly, the presence of these
charged particles changes the refractive index of the interstellar
medium, and therefore the group velocity of the radio waves. In the
context of pulsar timing, this is observed as a dispersion delay that
strongly scales with the observing frequency ($\tau_{DM}\propto
\nu^{-2}$) and which is characterised by the dispersion measure (DM),
defined as the integrated electron density between the pulsar and
Earth, in units of cm$^{-3}$pc \citep{lk05}.

This dispersive delay in itself does not affect high-precision pulsar
timing experiments, but because pulsars are high-velocity objects
\citep{hlk+04}, the integrated electron density between Earth and a
given pulsar changes slightly as the line of sight samples different
regions of the turbulent interstellar medium (ISM)
\citep{bmg+10}. This makes the DM and its associated
frequency-dependent delay time-variable, which in turn introduces
unmodelled variations in the pulsar timing data.  This time
variability is the main cause of unmodelled ToA variations in
high-precision pulsar timing data \citep{vbc+09}; therefore,
mitigating this effect is a critical challenge in moving towards
gravitational-wave detection \citep{kcs+13,lbj+14}.

There are essentially three approaches to dealing with this problem;
and most likely a combination of approaches will be needed in order to
obtain the best results.

Firstly, because the dispersive ISM delay is strongly
frequency-dependent, observations with large bandwidths can measure
these delays with very high accuracy.  \cite{lbj+14} have evaluated
the covariance of DM variation and timing precision, where they found
that a wider band and better precision is necessary to reduce such
covariance.  Several pathfinder projects where ultra-broadband
receivers simultaneously observe at all frequencies between
$\sim$500MHz and $\sim$2 GHz are ongoing, and the efficacy of this
approach is being evaluated.  This method is unique because the data
used for the high-precision timing project are the same data used for
DM determination, allowing a simultaneous determination of all
parameters at the cost of increased covariance between parameters.
The main challenge of this method involves possible corruptions
introduced by unmodelled frequency-dependence of the pulse profile and
possible frequency-dependent behaviour of pulse phase jitter (see
\S\ref{sec:jitter}).  In recent work by \cite{ldc+14} and \cite{pdr14}
algorithms have been presented to model the frequency dependence of
the pulse profile and improve the timing accuracy of pulsar
observations obtained with wide band instruments.  However, to
accurately asses these problems SKA1 sensitivity is required.

A second approach is, instead of determining the DM values
simultaneously across a very wide bandwidth, to use only the very
lowest frequencies available ($\sim$100--300 MHz) to determine the DM
values independently from the timing data obtained at higher
frequencies.  This approach puts smaller demands on available
observing time and does not require a very high instantaneous
bandwidth. One potential problem with this approach, however, is that
the lowest observing frequencies may actually sample a different part
of the ISM than the higher observing frequencies, because the
refractive angles are different for different frequency bands. In
practice this effect should be limited for a homogeneous and smooth
ISM, but the turbulent ISM can hardly be expected to be smooth and
homogeneous. This approach is currently being tested with the LOFAR
pathfinder and as yet the inhomogeneities of the ISM do not seem to
pose a problem \citep{hsh+12}.

Finally, with sufficient sensitivity, a large telescope could collect
pulsar timing data at high observing frequencies (3~GHz
and up). This would strongly diminish the effects of DM variations and
thereby avoid the need to correct for variable interstellar delays.
The steep spectral index of most pulsars will require a longer
integration time at these frequencies, but this may be required to
combat the effects of pulse phase jitter. The clear disadvantage of
this method, is two-fold: most importantly, without corrections for
variable DMs, the red noise spectrum of the DM variations
\citep{ars95} will bring these variable delays to a level where they
do affect the timing and hence gravitational-wave sensitivity. At that
point, if no other multi-frequency information is available the data
set will no longer be useful, as back-correction for DM variability is
not possible. A secondary draw-back is that the spin-off science of
studying the hot ionised component of the ISM would become impossible
with high-frequency data alone.

A more subtle effect of the ionised interstellar medium is
interstellar scattering. This effect is primarily observable through
relatively innocuous variations in the pulsar's brightness
(scintillation), caused through interference between rays with
slightly different path lengths. In a more extreme form, this
difference in path length between rays that are scattered off
different density inhomogeneities in the ISM actually causes
significant amounts of flux to arrive at the telescope with an
observable delay. This causes the shape of the pulse profile to
display an exponential decay (a so-called ``scattering tail''). This
scattering degrades timing precision because the scattering tail
washes out any fine-scale features of the pulse profile (which are
particularly useful in locking down the phase of the observed pulses).
More importantly, if the scattering is time-variable (and, again,
given the pulsar's high spatial velocity and the turbulent character
of the ISM, this is to be expected), then the shape of the pulse
profile will vary in time, affecting the timing in an unpredictable
manner. Because of the strong dependence of scattering on density
inhomogeneities, the scaling of scattering with frequency is less
well-defined than for dispersive delays, but scattering strength
typically scales even more strongly with frequency ($\propto
\nu^{-4.4}$, \citealt{bcc+04}), providing another impetus for
high-frequency timing observations.  While it is impossible to remove
scattering exactly as in the case of dispersive delays, one promising
technique termed cyclic spectoscopy has been demonstrated to retrieve
the intrinsic, unscattered pulse profiles for one highly scattered,
bright millisecond pulsar \citep{dem11,wdv13}. 
A wider study is required to assess how broadly
this method will be applicable to less scattered and weaker
PTA pulsars.

\section{A realistic PTA using the SKA}\label{sec:realistic}

In this section we consider a realistic PTA project carried out using
the SKA.  We consider the expected number of pulsars, the plausible
timing precision achievable, the time span for the experiment,
specific issues relating to the SKA design, and the other telescopes
available for PTA observations that will be operating at the same
time as the SKA1.

\subsection{Telescopes observing pulsars in the SKA1 era}

While the SKA1 is being commissioned, the existing radio telescopes are
expected to continue their PTA observations. Even with the increased
sensitivity that the SKA1 provides, combining SKA1 data with the long
time baselines of the existing PTAs improves the sensitivity for the
detection of gravitational waves. Furthermore, it is essential for
long-term pulsar timing experiments that a long overlap period is
included between two observing systems in order to characterise the
timing offsets between them. Finally, obtaining PTA observations with
multiple telescopes provides greater cadence and coverage of
observing frequencies.

As the SKA1 is commissioned it is likely that most of the current 100-m
class telescopes will still be observing pulsars along with the
Five-hundred-meter Aperture Spherical Telescope (FAST) and the
Xinjiang Qitai 110m Radio Telescope (QTT) in China. The Giant
Metrewave Radio Telescope (GMRT) in India may begin to play a larger
role, as its capability to simultaneously observe at 300 and 600\,MHz
is useful for DM corrections. LOFAR could also be useful for DM
montoring of Northern pulsars. The Murchison Widefield Array will
complement LOFAR observations in the Southern hemisphere and can be
used to monitor effects of the interstellar medium.  In addition, the
upcoming 100-m class CHIME (Canadian Hydrogen Intensity Mapping
Experiment) will make daily observations of Northern-sky pulsars at
frequencies between 400 and 800\,MHz, providing densely-sampled
records of DM and scattering changes toward the relevant MSPs.

At the time that SKA1 is commissioned the ASKAP COAST project is
likely to be underway.  This project aims to discover new pulsars
(many of which may become part of the SKA PTA sample) and to carry out
timing observations. The ASKAP sensitivity is relatively low and
therefore it is likely that the majority of high precision timing
being carried out in the Southern hemisphere at the time will be being
obtained with Parkes telescope and the MeerKAT array. The baseline design indicates that
the MeerKAT array will be combined as part of the \skamid.  However,
it will be essential either for the Parkes telescope to provide an
overlap period with the SKA1, or for MeerKAT to continue to operate in
an overlap mode with the \skamid\ for at least 1 year.

With the collecting area of FAST and the \skamid, it is likely that most
of the currently observed millisecond pulsars will be jitter
dominated.  An analysis of a PTA for the FAST telescope \citep{hdm+14}
showed that 1) long observation times $\sim 1$\,hour per pulsar will
still be required to obtain high time-precision observations for most
of the currently known pulsars and 2) for pulsars that are jitter
dominated there is no advantage (for timing purposes) in using a very
large telescope \citep[see also][]{lkl+12}.  However, the \skamid\ has the
advantage that sub-arrays can be formed enabling observations of
multiple pulsars (at, or close to, the jitter level) simultaneously.

Optimising the observation strategy has been investigated by
\cite{lbj+12}, where both the single and multiple telescope
optimisations were considered under the influence of radiometer,
jitter, and red noise. They found that, because of jitter noise, it is
preferable to use a smaller telescope to observe the brightest
pulsars. In this way, for certain bright pulsars, it can be more
efficient to split the \skamid\ (or \skalow) into sub-arrays or use existing 100-m class telescopes.

\subsection{Observing strategy}

Current predictions for the detectable GW signal suggest that high
cadence observing is important, but ensuring a long baseline of
observations over years to decades is essential. A realistic scenario
would be to observe a sample of pulsars once per week.  In order to
obtain the highest possible timing precision it will be necessary to
obtain high quality polarisation calibration. This will require
dedicated pointing directly at the pulsar.  It is unlikely that more
than one PTA-quality pulsar will be within the primary beam of the
telescope, meaning that multiple tied-array beams will not necessarily
be useful for the high precision timing programme, however the
remaining beams can be used in parallel for other timing projects.
The full sensitivity of the telescope will not be required in
cases where the pulsars are dominated by jitter noise. In that case,
it is more efficient to sub-array the telescope to observe multiple
pulsars simultaneously.

It is likely that around 50 pulsars
could be observed during a single 24-hour observing block providing a
timing precision of $<$100\,ns.  This gives observation times of $\sim
15$\,minutes per pulsar (assuming some nominal setup time for each
observation) assuming that each pulsar is observed with two
receivers. In order to account for DM and scattering variations it
will be necessary to observe at a wide frequency range.  Because of the
steep radio spectra of pulsars, current PTA observations are limited
to frequencies below $\sim 3$\,GHz. With \skamid, PTA observations
will be carried out with bands 2, 3 and 4 (0.95 to 1.76\,GHz, 1.65 to
3.05\,GHz and 2.80 to 5.18\,GHz respectively).  As discussed in
\S\ref{sec:dm}, lower frequency bands, and \skalow\ should be used for
tracking variations in dispersion measures. 

\begin{figure}
\includegraphics[width=.35\textwidth,angle=-90]{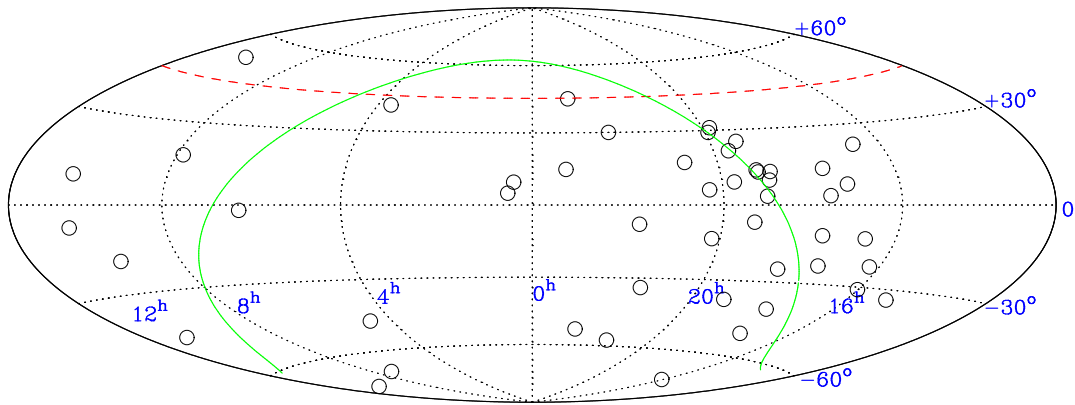}
\caption{Expected sky coverage of the SKA, overlaid with the positions
  of the current pulsars being observed as part of the IPTA project
  (open circles).  For an array mainly at latitude $-30^\circ$ the
  telescope will have an approximate declination range from
  $-90^\circ$ to $+45^\circ$ (red dashed line). We note that 49 out of
  the 50 current IPTA pulsars could be observed by the SKA.  The green
  line indicates the Galactic plane.}\label{fg:ska_sky}
\end{figure}

\subsection{Offline data analysis}

The data taken for the current PTA projects are generally stored in
greatly compressed form, with typical data file sizes for a single
pulsar observation currently around $\sim1$GB (for a data file
containing 64 subintegrations, 1024 frequency channels and 2048 phase
bins). These data sizes have not posed any major challenges for
current timing programmes.  However, as data rates increase, recording,
archiving and processing are all becoming more challenging.  With
wider bands available it is plausible that the SKA would produce data
files of $\sim10$\,GB for each standard observation. A year of
observing 100 pulsars with weekly cadence would lead to $\sim60$\,TB
for each backend instrument.  This is not impractical, but it will be
necessary to provide copies of these data at the data processing
centres. Furthermore, it is likely that new methods for high-precision
timing will soon be developed.  These could include using bright
individual pulses, higher-order moments of the electric field, or
dynamic spectra, all of which would require additional processing and
the storage of additional products.  Some methods, such as cyclic
spectroscopy, may even require the storage of all data in raw,
baseband form.

The supporting infrastructure that enables multiple copies of the data
to be replicated at different sites and enables the processing of that
data will need to be planned in detail.  Keeping these systems
operating and providing data and processing power to the community is
non-trivial and will require significant management.  Standard
processing of the pulsar arrival times requires minimal computing
resources.  A typical set of 260 observations for a given pulsar
spread over five years can easily be processed with the pulsar timing
software package \textsc{tempo2} \citep{hem06} on a laptop computer.
However, with the wide bandwidths available it is likely that a large
number of ToAs at different frequencies will be calculated and stored
for a single observation.  Many of the gravitational wave detection
codes (both frequentist and Bayesian based methods) require the
analysis of $N_{\rm toa} \times N_{\rm toa}$ matrices and many of the
algorithms require $N_{\rm toa}^3$ operations.  This can quickly
become prohibitive for a large number of ToAs, particularly if it is
necessary to also run the detection code on a large number of
simulated data sets.  It is therefore essential that the data sets can
be processed on high performance computers.  Much of the pulsar
processing software is now being modified to allow the use of Graphics
Processing Units (GPUs) where available.  GPUs are well suited for
dealing with processing large matrices and we therefore recommend that
the infrastructure created for the SKA allows the data to be processed
both on Central Processing Units (CPUs) and also on GPUs.

\subsection{Time to detection}\label{sec:cut}

\begin{figure}\centering
\includegraphics[width=.49\textwidth]{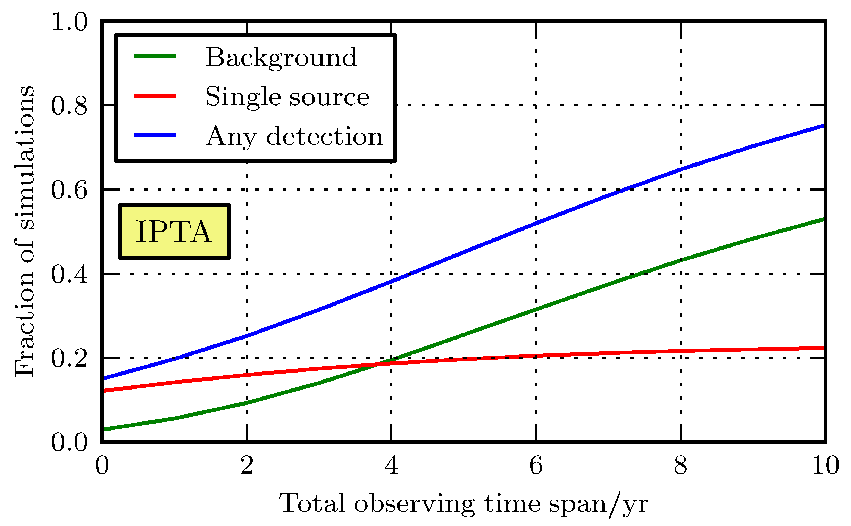} 
\includegraphics[width=.49\textwidth]{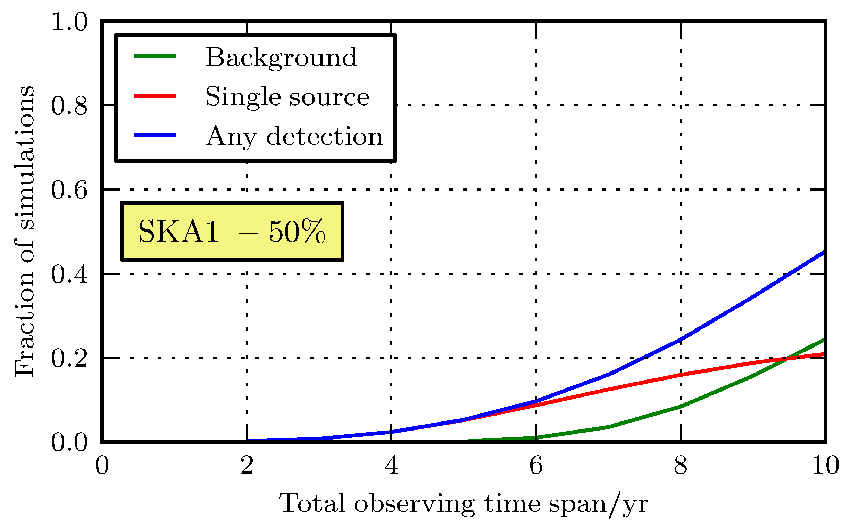} 
\includegraphics[width=.49\textwidth]{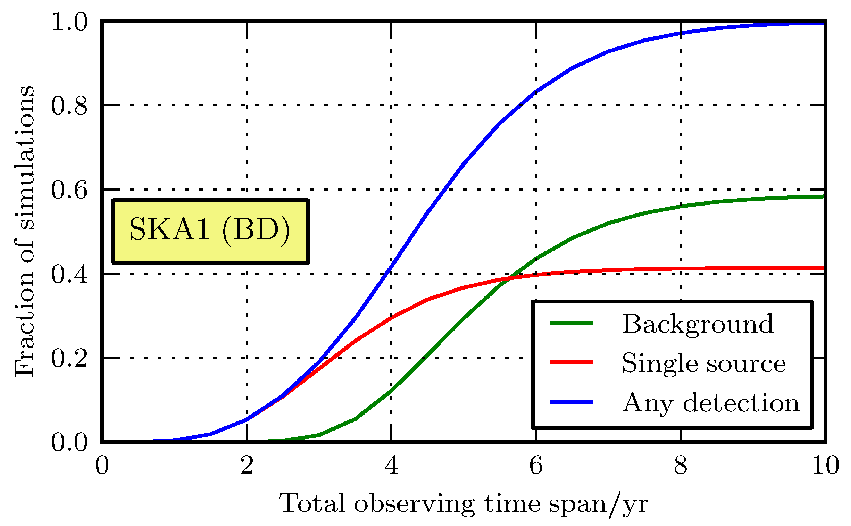} 
\includegraphics[width=.49\textwidth]{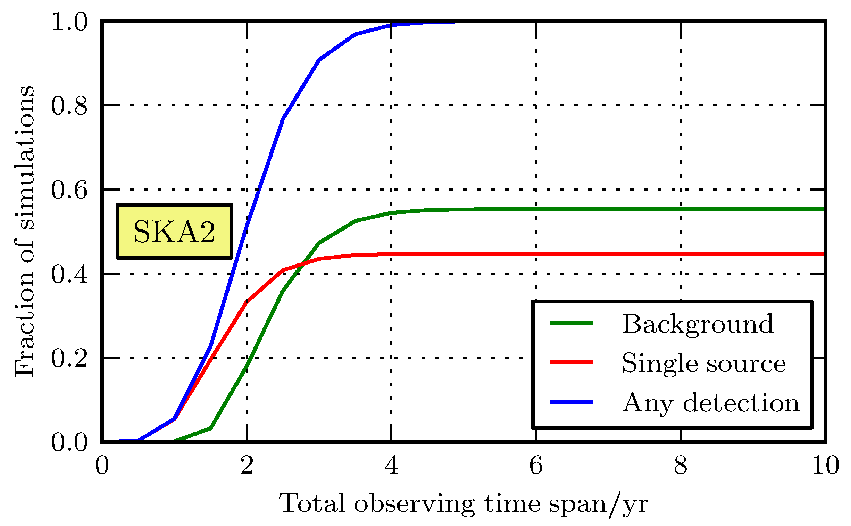} 
\includegraphics[width=.49\textwidth]{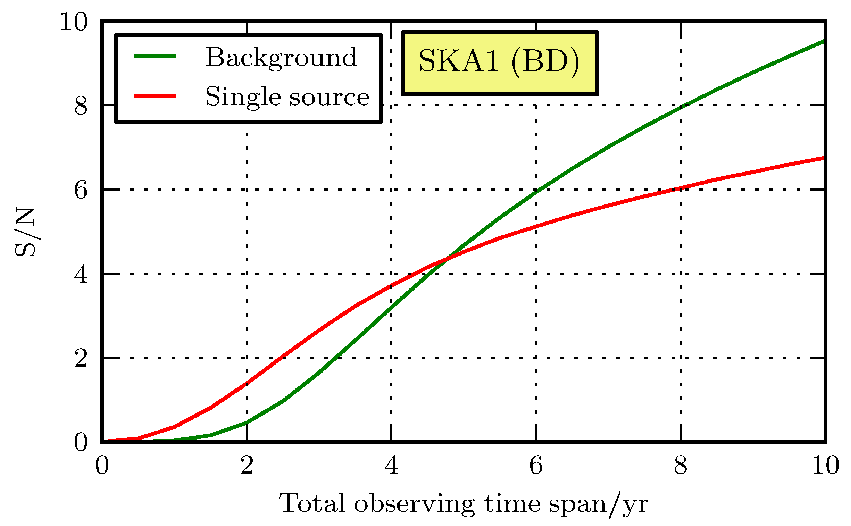}
\includegraphics[width=.49\textwidth]{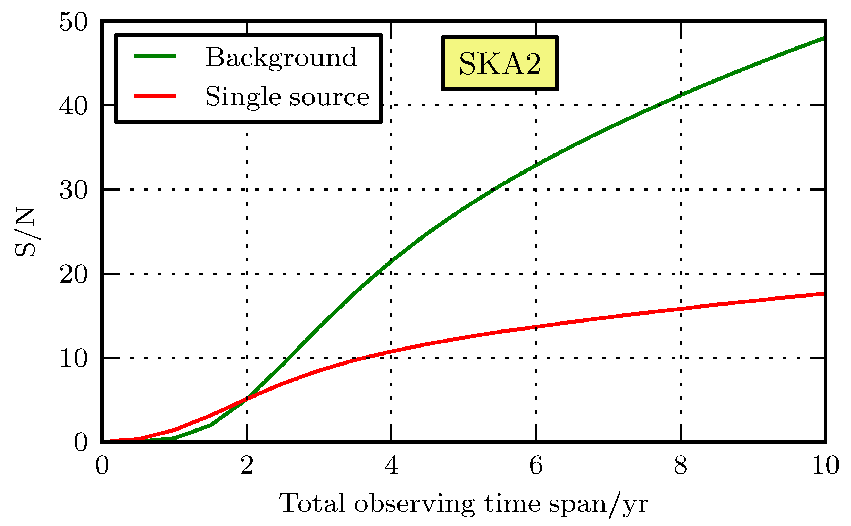}
\caption{Detection probability for four different PTAs; top left:
  IPTA; top right: SKA1 at an early science (50\%) sensitivity level;
  middle left: SKA1 as defined in the baseline design document; middle
  right: SKA2. The top and middle panels show the predictions on the
  \emph{first} GW detection by different PTAs as a function of
  time. In the IPTA case, the starting point corresponds to the
  present day, after 10 years of previous observations. In the SKA
  cases, the starting points correspond to the beginning of the SKA
  data-taking, disregarding previous data. The green curve gives the
  fraction of simulations of the Universe (which can be considered as
  a probability) in which a stochastic background is detected
  first. The red curve gives the probability of a single binary being
  detected first. The blue curve gives the overall probability of
  detecting any kind of GW signal.  We assume that a GW signal is
  detected when it produces a signal-to-noise ratio (S/N) larger than
  a certain threshold. The chosen S/N threshold is 4, which may be
  considered realistic for single binaries, but a bit conservative for
  a background. A more detailed study can be found in Rosado, Sesana
  \& Gair (in prep.). The S/N of a detection of a single binary and of
  a background are obtained as described in \cite{sv10} and
  \cite{sejr13}, respectively, and are plotted as function of time in
  the bottom plots for SKA1 and SKA2 PTAs. }\label{fg:timetodetect}
\end{figure}

In general, high-precision pulsar timing experiments rely on a sensitive telescope
with sufficient frequency coverage to mitigate noise sources.  
As mentioned in \S\ref{sec:studyGWs}, the success of detecting and
studying GWs with the SKA depends on the quality of the observations
and the number of pulsars in the array. In
Figure\,\ref{fg:timetodetect}, a comparison is shown between different
PTA configurations and the probabilities of detecting a GW background
or a continuous GW source.  To produce these plots, $\sim10^5$
simulations of the GW signal from SMBHBs have been analysed (Rosado,
Sesana \& Gair in prep.).  All the models assumed in the production of
these simulations are consistent with current electromagnetic
observations \citep{ses13b}.  

With the SKA as described in the baseline design document, it will
have a 50\% probability of detecting a GW source (either single or
background) after five years. This is a conservative estimate since
pre-existing IPTA data sets have not been included in the probability
calculations.
The PTA project on the SKA would not be significantly affected by a modification
to the maximum and minimum baselines available.  However, significant
changes to the GW-sensitivity of the PTA would occur if a major reduction
was made to the telescope sensitivity (e.g.  fewer dishes or degraded
receiver performance), if only a small frequency band was available or
without the possibility of sub-arraying the telescope.
The main effect of a reduced sensitivity on the SKA1 is a prolonged
timescale for all GW detections.  For example, in the case that the
sensitivity of the SKA1 is reduced by 50\%, the overall GW detection
probability after five years is reduced to 10\%.

\section{The future of gravitational wave astronomy}\label{sec:gwastronomy}

SKA1, or the IPTA using pre-existing telescopes, is likely to make the first
direct detection of GWs and will start to probe the properties of
those waves.  With the increased sensitivity of SKA2,
gravitational wave astronomy will become a reality and the detailed
properties of GWs can be studied.  Unexplored territories of both
astronomy and physics will be unveiled for the first time. The
foreseeable breakthroughs include:

\begin{itemize}
  \item \emph{Constraining models for galaxy evolution:} Determining the
    GW properties (amplitude, spectral index, anisotropy or source
    distribution) of a GW background from coalescing supermassive
    black hole systems will lead to an improved understanding of
    galaxy evolution and black hole growth. With SKA2 it will
    eventually be possible to study the high end of the SMBHB mass distribution at
    low redshifts.
\item \emph{Source localisation and population:} By decomposing the
  stochastic GW background on a basis of spherical harmonics
  \citep{msmv13}, it is possible to characterise the GW background on
  any angular scale up to the maximum angular resolution of the
  PTA. This characterisation of the background can confirm its
  cosmological origin, but is also completely general. It is therefore
  possible to use this method to find GW hotspots, which may help to
  confine sky areas which need to be covered by continuous wave
  searches. Bayesian search techniques have been recently developed by
  \cite{tg13}, and are being implemented in anisotropic stochastic GW
  background searches. A CMB-like search has also been recently
  proposed by \cite{grtm14}, which can be used to recover the
  components of an arbitrary background.
\item \emph{Testing gravity in the strong-field regime:} Although
  there is still lack of a consistent framework to quantise the well
  accepted theory of general relativity, the quantisation for
  weak GWs was succesfully initialised in the 1960s. There are two
  basic intrinsic properties for particles: their mass and spin
  states. In general relativity, the corresponding particles, the
  gravitons, are zero-mass spin-2 particles. However, other
  theories of gravity predict up to six polarisation states of
  GWs. Investigating the polarisation
  states and constraining the mass of the graviton with a PTA
  \citep{ljp08,ljp+10,cs12,cs12a,lee13} will help to understand the
  intrinsic symmetry of the gravitational interaction and may give
  valuable inputs for quantising gravity in the strong-field regime,
  which is one of the key questions of modern theoretical physics.
  While timing of binary pulsars is providing important insights into
  the dynamics of gravitation with strongly-gravitating bodies
  \citep{ss+14}, PTAs will present an unique opportunity to study
  gravitation in the radiative regime with GWs.
\item \emph{Studying and testing cosmological GWs:} As shown by
  \cite{sbs12}, cosmic strings are predicted to generate a different
  GW background compared to a background generated by SMBHBs.  The
  properties of a cosmic string-generated GW background are defined by
  their string linear energy density, the details of the string GW
  emission mechanism (kinks and cusps), the birth-scale of cosmic
  string loops, the intercommutation probability (the probability with
  which two cosmic string segments exchange partners when crossing
  each other) and possible deviations from the simple scaling
  evolution \citep[e.g.][]{sa13}. By identifying the GW background
  generated by cosmic strings, we will get access to a unique
  cosmological lab that will give us access to physics at an energy
  scale close to the GUT scale. In the case of cosmic superstrings,
  the expected results are even more profound since we will achieve a
  measurement of a fundamental string theory parameter for the first
  time, such as the fundamental string coupling or the
  compactification/warping scale. By measuring their properties, we
  will be able to investigate underlying dynamics and make a
  distinction between cosmic strings and superstrings.  Besides the
  cosmic string stochastic GWB, an SKA PTA project might also be able
  to detect a stochastic GWB created by several other mechanisms in the early Universe, \citep[e.g.][]{ffdg09,fhu13}.
\item \emph{GW distance scales:} When a sub-lightyear accuracy
  for pulsar distances can be achieved with SKA2 \citep{lwk+11},
  Galactic dynamics can be investigated. 
 A PTA is a unique type of GW detector working in the short wavelength
 regime, i.e. many radiation wavelengths separate pulsar pairs from
 each other and the Earth. This makes the angular resolution of PTA
 higher ($\delta \theta\simeq\lambda/D\sim1$\,arcmin; \citealt
 {lwk+11}) if we know the exact pulsar distance. With \skamid, the
 pulsar distance will not be measured accurate enough, which prevents
 us from achieving such high resolution. Nevertheless, anisotropies in
 a GW background will still result in a different decomposition of the
 Hellings \& Downs curve. When the nature of the anisotropy can be
 measured \citep{tg13}, the nature of the population of the GW sources
 can be studied \citep{msmv13}. With the SKA2, the pulsar timing
 parallax will bring the pulsar distance error down to a few light
 years. Such precise distance information will: \\
 I) make the PTA signal to single GW sources coherent, allowing
 perform GW interferometric imaging of the whole sky with angular
 resolutions of arcminutes \citep{lwk+11}. This will enable the search
 for electromagnetic counterparts to the GW sources, possibly enabling
 the follow-ups of GW sources with the traditional astronomical tools
 which can lead to the study of galaxy mergers \citep{bur13,rs14,dbc+14}.\\
 II) allow us to measure the mass and spin of a SMBHB and map the
 non-linear dynamics of the gravitational field \citep{mgs+12}; when
 the pulsar term can be distinguished from the Earth term, each pulsar
 term sees a slightly different piece of the evolutionary history of
 the binary since its orbital period evolves over the light travel
 time between the Earth and the pulsar. Concatenating these different
 snapshots of the waveform may allow for the measurement of the mass
 and spin of the binary, and furthermore, measuring post-Newtonian
 parameters up to 1.5 order.
\item \emph{Probing the inflationary era of the universe:} There are two
  major methods to probe the inflationary era of the universe with
  PTA. The first one is to examine relic GWs, which give constraints
  on the tensor-to-scalar ratio \citep{zzyz13}. The second one is to
  check the effects of structure formation, e.g. primordial black hole
  population, which give us information on the primordial density
  fluctuation \citep{sy09}.
\end{itemize}

\section{Conclusions}\label{sec:conclusion}
  
  The first direct detection of GWs in the nano-Hertz regime using PTA
  observations is expected within a decade. The SKA will play an
  important role in this, either through making the first detection,
  or by confirming the detection independently, and providing the
  sensitivity to characterise continuous sources as well as the
  stochastic GW background.  In general, pulsar timing precision will
  be greatly enhanced by the unprecedented collecting area and
  bandwidth of the SKA, and PTA sensitivity will be revolutionised
  though regular observations of a large number of pulsars. These
  would include many of the weaker pulsars wich may -- through the
  added sensitivity of the array -- for the first time be timed at the
  precision that is required for PTA work.

  In the early stages of the science, even before the SKA PTA achieves
  a direct GW detection, these data sets will already enable exciting
  scientific results; for example tests of theories of gravity,
  studies of binary and millisecond pulsars, searches for
  irregularities in the terrestrial time standards, and studies of the
  interstellar medium.  While these results are scientifically
  interesting in their own right, they will also aid the
  characterisation of the pulsars and the observing system in a way
  that continuously enhances our sensitivity to GWs.  Without a
  doubt, a PTA is an ideal experiment for the SKA.

\setlength{\bibsep}{0pt}

\end{document}